\documentclass[prb,a4paper,showpacs,groupedaddress,floatfix,superscriptaddress,twocolumn]{revtex4}
\usepackage[english]{babel}
\usepackage{amsmath,amssymb}

\usepackage{dsfont}
\usepackage{txfonts}
\usepackage{upgreek}
\usepackage{subfigure}
\usepackage{graphicx}
\usepackage{epstopdf}
\usepackage{dcolumn}

\usepackage{color}

\usepackage{float}
\usepackage{bm}

\usepackage{bm}
\usepackage{epstopdf}

\usepackage{hyperref}

\newcommand{\varH}{{\mathcal{H}}}

\newcommand{\mrm}[1]{\mathrm{#1}}

\newcommand{\sgn}{\mathop{\operator@font sgn}}

\newcommand{\ket}[1]{\left|#1\right\rangle}

\newcommand{\up}{\uparrow}
\newcommand{\down}{\downarrow}

\def\bra#1{\mathinner{\langle{#1}|}} 
\def\ket#1{\mathinner{|{#1}\rangle}} 
\newcommand{\braket}[2]{\langle #1|#2\rangle}

\newcommand{\new}[1]{\textcolor{black}{#1}}
\newcommand{\newred}[1]{\textcolor{black}{#1}}

\begin{document}

\preprint{APS/123-QED}

\title{Non-monotonic spin relaxation and decoherence in graphene quantum dots with spin-orbit interactions}

\author{Marco O. Hachiya}
\affiliation{Instituto de F\'{\i}sica de S\~ao Carlos, Universidade de S\~ao Paulo, 13560-970 S\~ao Carlos, S\~ao Paulo, Brazil}
\author{Guido Burkard}
\affiliation{Department of Physics, University of Konstanz, D-78457 Konstanz, Germany}
\author{J. Carlos Egues}
\affiliation{Instituto de F\'{\i}sica de S\~ao Carlos, Universidade de S\~ao Paulo, 13560-970 S\~ao Carlos, S\~ao Paulo, Brazil}

\date{\today}

\begin{abstract}

We investigate the spin relaxation and decoherence in a single-electron graphene quantum dot with Rashba and intrinsic spin-orbit interactions. 
We derive an effective spin-phonon Hamiltonian via the Schrieffer-Wolff transformation in order to calculate the spin relaxation time $T_1$ and decoherence time $T_2$ within the framework of the Bloch-Redfield theory.  
In this model, the emergence of a non-monotonic \new{dependence of} $T_1$ on the external magnetic field is attributed to the Rashba spin-orbit coupling-induced anticrossing of opposite spin states. 
A rapid \new{decrease} of $T_1$ occurs when \new{the spin 
and orbital relaxation rates become comparable} in the vicinity of the spin-mixing energy-level anticrossing. 
By contrast, the intrinsic spin-orbit interaction \new{leads to
a monotonic magnetic field dependence of the spin relaxation rate
which is caused solely by the direct spin-phonon coupling mechanism}.
Within our model, we demonstrate that the decoherence time $T_2 \simeq 2 T_1$ is dominated by relaxation processes for the electron-phonon coupling mechanisms in graphene up to leading order in the spin-orbit interaction. 
Moreover, we show that the energy anticrossing also leads to a vanishing pure spin dephasing rate for these states for a super-Ohmic bath. 
\end{abstract}

\pacs{Valid PACS appear here}
\maketitle

\section{\label{sec:intro} Introduction}

Carbon-based materials such as graphene and carbon nanotubes are of recognized importance for their potential spintronic and quantum computation applications. 
Notably, single-layer graphene, \new{a} one-carbon-atom-thick layer
arranged in a honeycomb crystal lattice, has 
\new{attracted} much interest in the last decade due to its unique electronic properties\cite{CastroNetoReview}. 
The electron spin degree of freedom in graphene quantum dots makes them promising candidates for universal scalable quantum computing \cite{Loss98, Guido07}, which would rely on spin relaxation and
decoherence times \new{much longer} than the gate operation times \cite{DiVincenzo}. 
Graphene has a relatively weak hyperfine interaction and spin-orbit (SO) couplings.
A graphene sheet is composed naturally of $99 \%$ of $^{12}C$ with
nuclear spin $0$, and \new{of} $1 \%$ $^{13}C$ with nuclear spin $1/2$, leading us to long dephasing times in carbon-based quantum dots due to a weak hyperfine interaction\cite{Fischer09}. 
Thus graphene \new{emerges} as a good candidate to host a spin qubit, in contrast to GaAs quantum dots, whose spin dynamics is strongly modified by the nuclear spin bath. 
Moreover, the weak SO couplings in graphene generates \new{a} spin-splitting on the order of tens of $\mu eV$ due to the low atomic weight of carbon atoms\cite{KaneMele05, Min06}. 
Long spin relaxation times are expected since the mechanisms that enable relaxation channels arise as a combined effect of
\new{non-piezo-electric} electron-phonon interaction and \new{ weak} SO coupling.

Despite the lack of measurements of the spin relaxation and dephasing
times in graphene quantum dots, experimental results have already been
reported in a two-electron $^{13}C$ nanotube double quantum dot\cite{Churchill09} that has been \new{isotopically-enriched}. 
These results showed a non-monotonic magnetic field dependence of the spin relaxation time near the energy anticrossing. In this case, the spin relaxation minimum is related to the coupling between electron spin in the quantum dot and the nanotube deflection\new{\cite{Bulaev08,Rudner10}}.

\begin{figure}[htb]
\includegraphics[width=.5\textwidth]{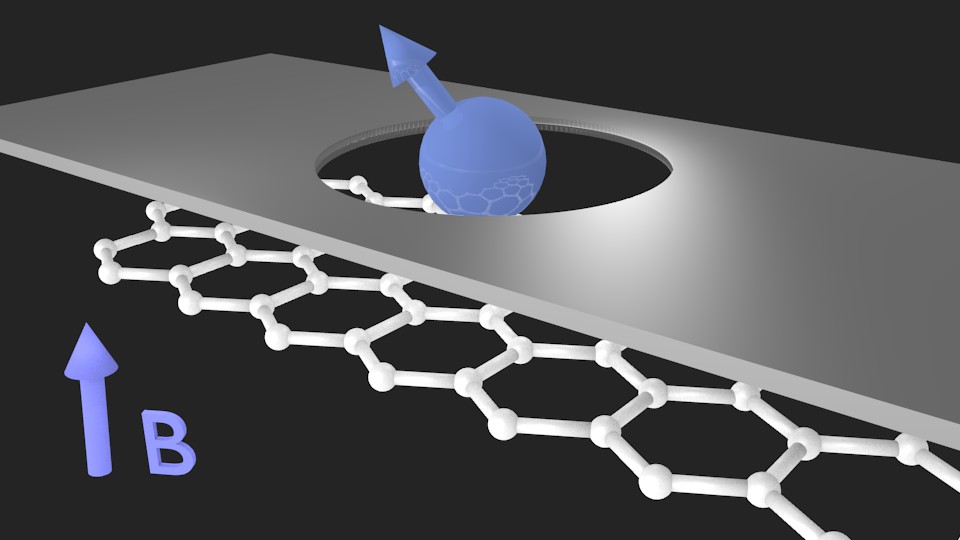}
\caption{\label{fig:setupgraphene} Schematic of a gate-tunable circular graphene quantum dot setup.
An homogeneous magnetic field is applied perpendicularly to the gapped graphene sheet.
A metallic gate put on top of the graphene defines the confinement potential for a single-electron.
Figure not drawn to scale.}
\end{figure}

In this paper, we derive a spin-phonon Hamiltonian using the Schrieffer-Wolff transformation for all mechanisms of electron-phonon and spin-orbit interactions. This effective Hamiltonian captures the combined effect of the SO interaction and electron-phonon-induced potential fluctuations.
Within the Bloch-Redfield theory, we find that a non-monotonic behavior of the spin relaxation time \new{occurs} as a function of the external magnetic field around the spin mixing energy-level anticrossing by the Rashba SO coupling in combination with the deformation potential and bond-length change electron-phonon mechanisms.
We predict that the mininum of the spin relaxation time $T_1$ could be experimentally observed in graphene quantum dots. 
This energy anticrossing takes place between the first two excited energy levels at the accidental degeneracy for a certain value $B^{*}$ of the external magnetic field.
We treat the accidental degeneracy mixed by the Rashba SO coupling using degenerate-state perturbation theory.
$T_1$ strongly increases at the energy anticrossing, reaching the same order as the orbital relaxation time\cite{Bulaev05, Stano05, Stano06}.  
In contrast with carbon nanotubes, the intrinsic SO does not couple these states due to the selection rules in a circular quantum dot, exhibiting a monotonic magnetic field dependence of $T_1$ due to direct spin-phonon coupling (deflection coupling mechanism). 
We also demonstrate that pure spin dephasing rates vanish in the leading order of the electron-phonon interaction and SO interactions causing a decoherence dominated by relaxation processes, i.e. $T_2 = 2 T_1$.
Moreover, we find a vanishing spin dephasing rate for a super-Ohmic \new{bath} as a general property of the energy anticrossing spectrum. 
This paper is organized as follows: In Sec.~\ref{sec:model}, we introduce the model to describe a circular graphene quantum dot.
In Sec.~\ref{sec:effsph}, we derive the effective spin-phonon Hamiltonian.
In Sec.~\ref{sec:spinrelaxationrates}, we present a calculation of the spin relaxation time $T_1$ within the Bloch-Redfield theory. 
In Sec.~\ref{sec:spindephasingrates}, we discuss the vanishing spin dephasing rate within our model.
Finally, we summarize our \new{results and draw our conclusions} in Sec.~\ref{sec:conclusion}.

\section{\label{sec:model} The Model}

In this section, we introduce the model for a circular and gate-tunable graphene quantum dot. 
Within our model, we consider a gapped graphene taking into account electron-phonon coupling mechanisms and spin-orbit interactions.
We also analyze the energy spectrum of the quantum dot and its energy-level degeneracy. 
The degenerate levels are mixed by the Rashba SO coupling, and the energy crossings are removed using the standard degenerate perturbation theory. 

\subsection{\label{sec:gdots} Graphene quantum dots}

The low-energy effective Hamiltonian for graphene is analogous to the
two-dimensional massless Dirac equation. The characteristic linear
dispersion for massless fermions occurs at the \new{two}
non-equivalent points $K$ and $K^{\prime}$ \new{(valleys)}, in the honeycomb lattice Brillouin zone.
The graphene energy bands in the vicinity of these high-symmetry
points constitute a \new{solid-state realization of} relativistic quantum mechanics. 
However, confining electrons in graphene
quantum dots \new{is} a difficult task, 
since \new{the particles} tend to escape from the electrostatic confinement potential due to Klein tunneling. 
This problem can be overcome \new{by} putting graphene on top of a substrate, such
as SiC\cite{Zhou07} and BN\cite{Gio07, Sla10}, that induces a
non-equivalent potential for each atom of the two carbon sublattices and adds a
mass term to the Hamiltonian \new{\cite{Recher09}}.
The sublattice A(B) will feel a potential parametrized by $+(-) \Delta$ which breaks inversion symmetry, opening a gap $2 \Delta$ in the electron-hole energy spectrum.
Combined with the mass term, an external magnetic field $\mathbf{B}$ is necessary to break the time-reversal symmetry and lift the valley degeneracy. 
Thus it is reasonable to confine a single electron in a quantum dot with the restriction of its being localized in a single valley.
Consider then, a circular and gate-tunable graphene quantum dot in an
external magnetic field with SO interactions and the electron-phonon
interaction described by the following low-energy Hamiltonian for
the $K$ valley \new{\cite{Recher09}},
\begin{equation}
\label{eq:HQD1}
\varH= \varH_{\mrm{d}} + \varH_{\mrm{Z}} + \varH_{\mrm{SO}} + \varH_{\mrm{ph}} +\varH_{\mrm{e-ph}},
\end{equation}
with the quantum dot Hamiltonian $\varH_d$ and the Zeeman term $\varH_Z$, respectively, given by
\begin{equation}
\label{eq:HQD2}
\varH_{\mrm{d}}=\hbar v_{\mrm{F}} \mathbf{\Pi} \cdot \boldsymbol\sigma
+U(r) +\Delta\sigma_{z}, \quad \varH_{\mrm{Z}} = \frac{1}{2}g\mu_{B}\mathbf{B}\cdot\mathbf{s},
\end{equation}
where $\mathbf{\Pi}=\mathbf{p}-e\mathbf{A}$ is the canonical momentum. The vector potential is chosen such that  $\mathbf{B}=\nabla\times\mathbf{A}=(0,0,B)$, i.e., perpendicular to the graphene sheet. 
Here, $v_{F}=10^6 \mrm{m/s}$ is the Fermi velocity, $U(r)=U_0 \Theta(r-R)$ is the circular-shaped electrostatic potential, with $\Theta(x)=1$ for $x \geq 0$ and $\Theta(x)=0$ for $x < 0$. 
The operator $\boldsymbol\sigma$ acts on the pseudospin subspace (A,B sublattices), \new{while} $\mathbf{s}$ acts on the real spin. Both operators $\boldsymbol\sigma$ and $\mathbf{s}$ are represented by Pauli matrices. 
The SO Hamiltonian for the $K$ valley reads\cite{tensor}
\begin{equation}
	\varH_{\mrm{SO}} =
	\varH_{\mrm{i}}+\varH_{\mrm{R}} = 	\lambda_{\mrm{i}}\sigma_{z}s_{z} + \lambda_{\mrm{R}}(\sigma_{x}s_{y}-\sigma_{y}s_{x}),
\label{HSO}
\end{equation}
where $\varH_{\mrm{i}}$ and $\varH_{\mrm{R}}$ denote the intrinsic and
Rashba SO effective Hamiltonians\cite{KaneMele05}, respectively.
The intrinsic SO coupling \new{originates} from the local atomic SO interaction. 
At first, only the contribution from the $\sigma - \pi$ orbital
coupling \new{was considered}, resulting in a second-order term to the intrinsic SO coupling strength $\lambda_{\mrm{i}}$\cite{Min06}.
However, some $d$ orbitals hybridize with $p_z$ forming a $\pi$-band that gives a first-order contribution which plays a major role in the spin-orbit-induced gap\cite{Gmitra09}.
The Rashba SO coupling, also called the extrinsic contribution, arises when an electric field is applied perpendicular to the graphene sheet. 
The major contribution of the SO coupling $\lambda_{\mrm{R}}$ comes from the $\sigma-\pi$ hybridization\cite{Min06}, in contrast with the intrinsic case.
The Rashba SO could also be enhanced by curvature effects in the graphene sheet \cite{Huertas06}. 
\new{The free phonon Hamiltonian is given by}
\begin{equation}
	\varH_{\mrm{ph}} =  \sum_{{\bf q},\mu} \hbar \omega_{{\bf q},\mu} b^\dagger_{{\bf q},\mu}b_{{\bf q},\mu}
\label{Hph}
\end{equation}
with the dispersion relation \new{$\omega_{{\bf q},\mu} = s_{\mu} |{\bf q}|^{m}$}, where $s_{\mu}$ is the sound velocity and $m=1,2$ depending on the type of phonon branch. 
\new{Finally}, we have the electron-phonon interaction $\varH_{\mrm{e-ph}}$. We consider long-wavelength acoustic phonons represented by two main mechanisms: the deformation potential and the bond-length change mechanism\cite{Ando05}. 
The former is an effective potential generated by static distortions of the lattice. It is represented in the sublattice space as a diagonal energy shift in the band structure. 
The latter are off-diagonal terms due to modifications of the bond-length between neighboring carbon atoms, which causes changes in the hopping amplitude.   
%
\new{The electron-phonon interaction in the sublattice space is given by\cite{Ando05}}

\begin{equation}
\varH_{\mrm{e-ph}}=  \sum_{q, \mu} \frac{q}{\sqrt{A \rho \omega_{q,\mu}}} \left(\begin{array}{cc} g_1 a_1 & g_2 a_2^* \\ g_2 a_2 & g_1 a_1 \end{array} \right) (e^{i {\bf q} {\bf r}} b^{\dagger}_{q, \mu} -  e^{-i {\bf q} {\bf r}} b_{q, \mu}   ),
\label{eq:heph}
\end{equation}
where $g_1$ and $g_2$ \new{are the deformation potential and bond-length change coupling constants}.
Here, $A$ is the area of the graphene layer and $\rho$ is the mass area density. The constants $a_1$, $a_2$ and the sound velocities $s_{\mrm{LA}}$, $s_{\mrm{TA}}$ for the longitudinal-acoustic ($\mu=\mrm{LA}$) and transverse-acoustic ($\mu=\mrm{TA}$) modes are given in Table~\ref{tab:table1}.  
Both phonon branches have a linear dispersion relation given by $\omega_{{\bf q},\mu} = s_{\mu} |{\bf q}|$.
\new{Optical phonons} are not taken into account in this work, since their energies do not match the Zeeman splitting for typical laboratory fields. 
The out-of-plane phonons ($\mu=\mrm{ZA}$) will be discussed further below.
Notice that the electron-phonon interaction is spin independent and can only cause a spin relaxation when assisted by the SO interaction.  
\begin{table}[b]
\caption{\label{tab:table1}
Electron-phonon constants and sound velocities for longitudinal (LA) and \new{transverse} (TA) acoustic phonons.  
The phonon emission angle is denoted by $\phi_q$.
}
\begin{ruledtabular}
\begin{tabular}{cccc}
 & $a_1$ & $a_2$    & $s_{\mu} (10^4 \, {\rm m/s})$ \\
\hline
LA& $i$ & $i e^{2 i \phi_q}$ & 1.95\footnotemark[1] \\
TA& 0 &  $e^{2 i \phi_q}$ & 1.22\footnotemark[1]\\
\end{tabular}
\end{ruledtabular}
\footnotetext[1]{From Ref.~\onlinecite{Falko08}.}
\end{table}

In the following subsection, we analyze the bare quantum dot spectrum and perform a perturbation theory calculation for degenerate levels treating the SO Hamiltonian as a perturbative term.   

\begin{figure}[htb]
\includegraphics[width=.45\textwidth]{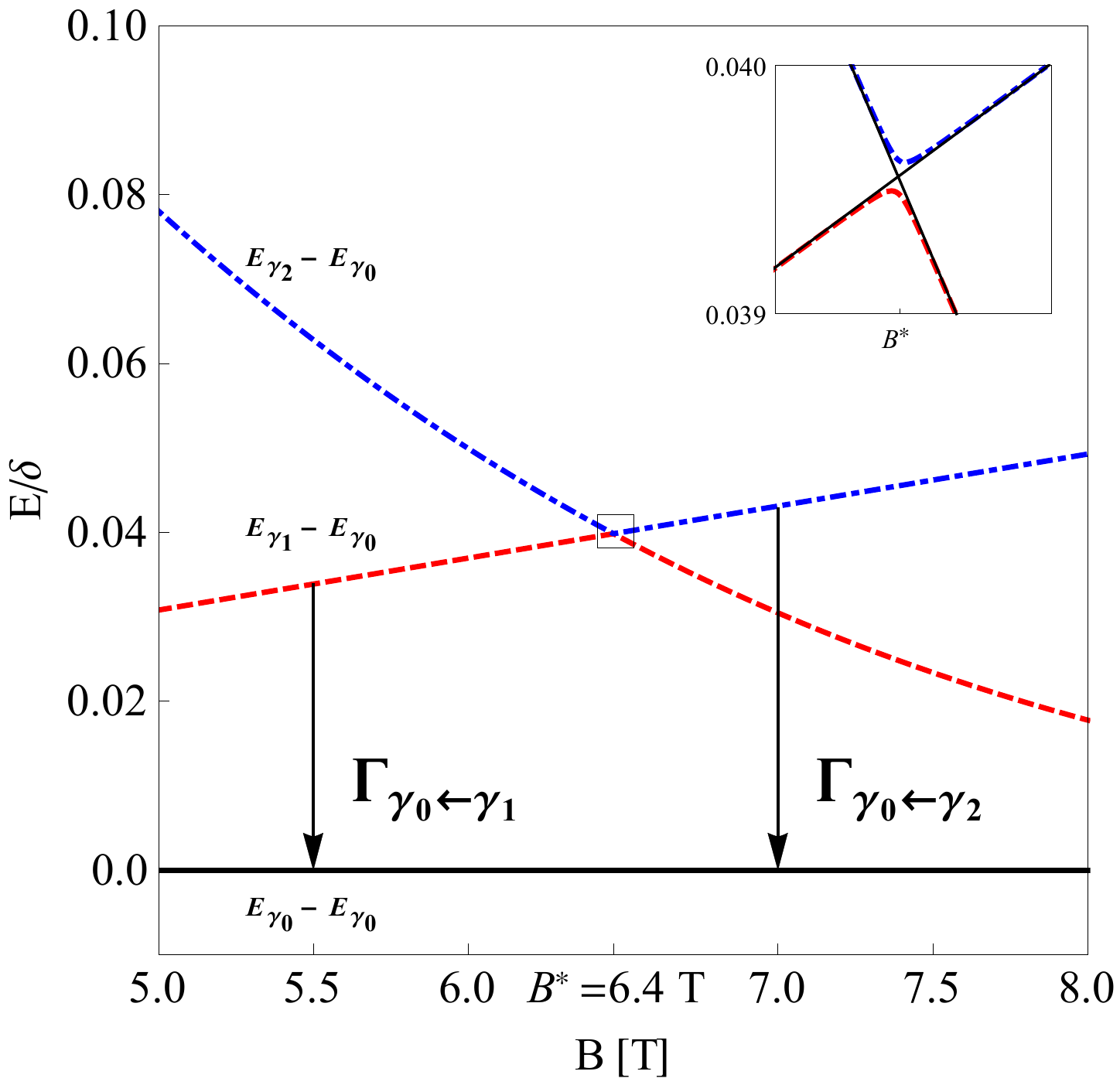}
\caption{\label{fig:plot1graphene} Magnetic field dependence of the energy difference between the perturbed three lowest energy levels and the ground state in a circular graphene quantum dot. 
\newred{Our spin qubit is composed by the ground state and the first excited state with opposite spin orientation.}
Sequentially from bottom to top, $E_{\gamma_0} - E_{\gamma_0}$ (solid), $E_{\gamma_1} - E_{\gamma_0}$ (dashed) and $E_{\gamma_2} - E_{\gamma_0}$ (dot-dashed).  
The Rashba SO interaction-induced anticrossing of the bare quantum dot states $E_{1/2,1,\uparrow}$ and $E_{-1/2,1,\downarrow}$, at $B=B^{*}$ (solid lines in the inset).
The spin relaxation rate takes place between the states $\ket{\gamma_0}$ and $\ket{\gamma_1}$ ($\Gamma_{\down \up} = \Gamma_{\gamma_0 \leftarrow \gamma_1}$) before the anticrossing, and between the states $\ket{\gamma_0}$ and $\ket{\gamma_2}$ ($\Gamma_{\down \up} = \Gamma_{\gamma_0 \leftarrow \gamma_2}$) after the anticrossing.
Inset: Blowup of the energy levels in the vicinity of the crossing region.}
\end{figure}

\subsection{\label{sec:deg} Degenerate state perturbation theory}

In order to calculate $T_1$ and $T_2$, we use the quantum dot eigenstates perturbed by the SO interaction. Before doing so, we have to get rid of the degeneracies in the quantum dot spectrum by applying degenerate state perturbation theory.
\newred{This procedure makes it clearer to define which states constitute our spin qubit and where the spin relaxation occurs.}
Due to the selection rules for the matrix elements of the SO interaction\cite{Struck10}, only the Rashba SO term couples states from the degenerate subspace.
Thus we intend to find a linear combination of eigenstates from the degenerate subspace of the quantum dot such that these states are not coupled by the Rashba SO Hamiltonian $\varH_{\mrm{R}}$. 
Consider then, first the bare quantum dot Hamiltonian in the $K$ valley
$\varH_{\mrm{d}}$, with $\varH_{\mrm{d}} \ket{\mrm{j,\nu,s}} = E_{\mrm{j,\nu}}\ket{\mrm{j,\nu,s}}$
and the quantum dot wave functions\new{\cite{Recher09}}
\begin{equation}
  \braket{r,\phi}{\mrm{j,\nu,s}}=
	\psi_{\mrm{j,\nu,s}}(r,\phi) = 
	e^{i(j-1/2)\phi}
	\begin{pmatrix}\chi_{A}^{\mrm{j,\nu,s}}(r) \\ 
		\chi_{B}^{\mrm{j,\nu,s}}(r)e^{i\phi}\end{pmatrix}.
		\label{solution}
\end{equation}
The spinor components $\chi_{\mrm{A,B}}^{\mrm{j,\nu,s}}(r)$ are proportional to the
confluent hypergeometric functions and are described by the set $
{\mrm{j,\nu},\mrm{s}}$, where we introduce the angular \new{($j = \pm 1/2, \pm 3/2,\ldots$),
radial ($\nu =1,2,3,\dots$) and spin $s = \up, \down$ quantum numbers}.
Matching the spinors at $r = R$ results in a \new{transcendental} equation for the eigenvalues $E_{\mrm{j,\nu}}$ which can be obtained numerically \cite{Recher09}. 
Since we are going to calculate the spin relaxation rates due to transitions between the lowest three energy levels of the quantum dot, we restrict ourselvels to the analysis of the subspace $\left\{ \ket{+1/2, 1,\down}, \ket{1/2,1,\uparrow}, \ket{-1/2,1,\downarrow}   \right\}$.
Including the Zeeman spin-splitting, it leads to a crossing of the energy levels $E_{1/2,1,\uparrow}$ and $E_{-1/2,1,\downarrow}$, for a certain magnetic field $\mathbf{B}^{*}$ depending on the size of the quantum dot.  
The ground state $\ket{+1/2, 1, \down}$ is not degenerate for any value of $\mathbf{B}$.
\newred{The Rashba SO interaction $\varH_{\mrm{R}}$ couples two of these states $\ket{+1/2, 1,\up}$ and $\ket{-1/2, 1,\down}$} due to its selection rule for the angular quantum number $j$\cite{Struck10}, which is given by $\left| j - j^{\prime}   \right| = 1$.
By contrast, the intrinsic SO interaction $\varH_{\mrm{i}}$ does not couple them since its selection rule is $\left| j - j^{\prime}   \right| = 0$. 
Now, we have to find an appropriate linear combination of the states from the \newred{degenerate
subspace  $ \ket{+1/2, 1,\up}, \ket{-1/2, 1,\down}$} \new{in which $\varH_{\mrm{R}}$ becomes diagonal}
in order to remove the accidental energy level degeneracy from the denominator in the usual non degenerate perturbation theory. 
Then, performing standard degenerate state perturbation theory, we obtain the zero-order eigenstates for the three lowest energy levels are given by	
\begingroup
\renewcommand{\arraystretch}{1.8}
\begin{equation}
  \begin{bmatrix}
   \ket{\gamma_0}
    \\
    \ket{\gamma_1}
    \\
    \ket{\gamma_2}
  \end{bmatrix}
  =
  \begin{bmatrix}
    1 & 0
    & 0
    \\
    0 & \displaystyle \cos(\vartheta/2) e^{i \delta}
    & \displaystyle -	\sin(\vartheta/2)
    \\
    0 & \displaystyle \sin(\vartheta/2) e^{i \delta}
    & \displaystyle \cos(\vartheta/2)
  \end{bmatrix}
  \begin{bmatrix}
    \ket{1/2,1,{\down}}  
    \\
    \ket{1/2,1,{\up}}
    \\
    \ket{-1/2,1,{\down}}
  \end{bmatrix},
\end{equation}
\endgroup   \\ 
with the associated first-order eigenvalues 

\begin{eqnarray}
E_{\gamma_0}=E_{1/2,1}- \frac{\hbar \omega_Z}{2},   & 
E_{\gamma_1, \gamma_2} = \epsilon_{+}  \mp \sqrt{\epsilon_{-}^2 + \left| \Delta_{\mrm{SO}} \right|^2},
\end{eqnarray}
plotted in Fig.~\ref{fig:plot1graphene}. We define $\epsilon_{+}=(E_{1/2,1}+E_{-1/2,1})/2$ and $\epsilon_{-}=(E_{1/2,1}-E_{-1/2,1} + \hbar \omega_Z)/2$, $\hbar \omega_Z = g \mu_B B$ is the Zeeman energy splitting.
Here, $\Delta_{\mrm{SO}}=\bra{1/2, 1,\up} \varH_{\mrm{R}} \ket{-1/2, 1, \down} = 4 \pi i \lambda_{\mrm{R}} \int \mathrm{dr} \; r \; \chi^{1/2,1}_{\mrm{A}}(r) \chi^{-1/2,1}_{\mrm{B}}(r)$, $\tan \vartheta	= \Delta_{\mrm{SO}}/ \epsilon_{-}$ and $\tan \delta	= \mathfrak{I}[\Delta_{\mrm{SO}}]/ \mathfrak{R}[\Delta_{\mrm{SO}}]$, where $\mathfrak{I}[x]$ is the imaginary part and $\mathfrak{R}[x]$ the real part of $x$. 
As a result, the Rashba SO induces an energy gap $2 \Delta_{\mrm{SO}}$ at the energy anticrossing ($\epsilon_{-} = 0$), as shown in Fig.~\ref{fig:plot1graphene}. We have two dominant spin components for $\ket{\gamma_1}$ and $\ket{\gamma_2}$ depending on whether the spin relaxation takes place before or after the energy anticrossing region. 
Before the energy anticrossing $\Delta_{\mrm{SO}}/ \epsilon_{-} > 0$, $\ket{\gamma_1} \approx \ket{1/2,1,{\up}} + \mathcal{O}( \Delta_{\mrm{SO}}/ \epsilon_{-})$ and $\ket{\gamma_2} \approx \ket{-1/2,1,{\down}} + \mathcal{O}( \Delta_{\mrm{SO}}/ \epsilon_{-})$.
Increasing the magnetic field we go through the energy anticrossing region such that $\vartheta \rightarrow \pi/2$ when $\epsilon_{-} = 0$. As a result, the states from the degenerate subspace hybridize $\ket{\gamma_1} \approx  \left(   \ket{1/2,1,{\up}} -  \ket{-1/2,1,{\down}}  \right)  /\sqrt{2}$ and $\ket{\gamma_2} \approx  \left(  \ket{1/2,1,{\up}} +  \ket{-1/2,1,{\down}}  \right)  /\sqrt{2}$.
After the energy anticrossing $\Delta_{\mrm{SO}}/ \epsilon_{-} < 0$, $\ket{\gamma_1} \approx \ket{-1/2,1,{\down}} + \mathcal{O}( \Delta_{\mrm{SO}}/ \epsilon_{-})$ and $\ket{\gamma_2} \approx \ket{1/2,1,{\up}}  + \mathcal{O}( \Delta_{\mrm{SO}}/ \epsilon_{-})$. 
Thus before the energy anticrossing, the spin relaxation takes place between $\ket{\gamma_1} \rightarrow \ket{\gamma_0}$ and after the energy anticrossing between $\ket{\gamma_2} \rightarrow \ket{\gamma_0}$. At the energy anticrossing, the spin up and down are equivalently mixed and the orbital relaxation rate dominates over the spin relaxation rate, since the latter is a higher-order process assisted by the SO interaction\cite{Bulaev05, Stano05, Stano06}. 
These results will be used to study the energy relaxation with spin-flip between excited states and the ground state.

\section{\label{sec:effsph} Effective spin-phonon Hamiltonian}

The electron-phonon coupling allows for energy relaxation between the Zeeman levels via the admixed states with opposite spin due to the presence of the SO interaction. 
To study this admixture mechanism we derive an effective Hamiltonian describing the coupling of spin to potential fluctuations generated by the electron-phonon coupling. 
We perform a Schrieffer-Wolff transformation in order to eliminate the SO interaction in leading order\cite{winklersbook, Golovach04},  
 
\begin{equation}
\widetilde{\varH} = e^{\mathcal{S}} \varH e^{\mathcal{-S}} = \varH_{\mrm{d}} + \varH_{\mrm{Z}} + \varH_{\mrm{ph}} +\varH_{\mrm{e-ph}} + \left[ \mathcal{S},  \varH_{\mrm{e-ph}}  \right]  ,
\label{eq:sw1}
\end{equation} 
where we have retained terms up to $\mathcal{O}\left( \varH_{\mrm{SO}}  \right)$\cite{sphonon}. 
The operator $\mathcal{S}$ obeys the commutator $\left[ \varH_{\mrm{d}} + \varH_{\mrm{Z}} , \mathcal{S}  \right] = \varH_{\mrm{SO}} $, with $\mathcal{S} \sim  \mathcal{O} \left( \varH_{\mrm{SO}}  \right) $.
The term $\left[ \mathcal{S}, \varH_{e-ph}  \right]$ represents the coupling of the electron spin to the charge fluctuations induced by the electron-phonon interaction via the SO interaction (admixture mechanism). 
The operator $\mathcal{S}$ can be rewritten as $\mathcal{S} = \left( L_{\mrm{d}} +  L_{\mrm{Z}}  \right)^{-1} \varH_{\mrm{SO}}$ where $\hat{L}_i$ is the Liouvillian superoperator defined as $L_{\mrm{i}} A = \left[ \varH_{\mrm{i}}, A \right]$, where A denotes an arbitrary operator.  
Here, we make the distinction $\mathcal{S} = \mathcal{S}^{\mrm{R}} + \mathcal{S}^{\mrm{i}}$, where $\mathcal{S}^{\mrm{i}} \propto \lambda_{\mrm{i}}$ and $\mathcal{S}^{\mrm{R}} \propto \lambda_{\mrm{R}}$. 
For the Rashba SO coupling, we have to consider the new basis $\left\{ \ket{\gamma_1}, \ket{\gamma_2} \right\}$ calculated in Sec.~\ref{sec:deg} using perturbation theory for the degenerate levels.  
As explained in Sec.~\ref{sec:deg}, we are interested in transitions from the excited states $\ket{\gamma_k}$ to the ground state $\ket{\gamma_0}$.
In this case, we calculate the matrix element of the effective spin-phonon Hamiltonian $\bra{\gamma_0} \varH_{s-ph}^R \ket{\gamma_k} = \bra{\gamma_0} \varH_{\mrm{e-ph}} + \left[ \mathcal{S}^{\mrm{R}},  \varH_{\mrm{e-ph}}  \right] \ket{\gamma_k} $, where $\gamma_k=\gamma_1, \gamma_2$. 
We find that

\begin{eqnarray}
\label{eq:hsphrashba}
\bra{\gamma_0} \varH_{\mrm{s-ph}}^R \ket{\gamma_k} &=& \bra{\gamma_0} \varH_{\mrm{e-ph}}  \ket{\gamma_k} \\ \nonumber
			&+& \sum_{\mrm{n,s} \neq \gamma_0} \frac{\Omega_1(\gamma_0, n, \gamma_k)}{E_{\gamma_0} - E_{n}}  \\  \nonumber
			&+& \sum_{\mrm{n,s} \neq \mathfrak{D}}   \frac{\Omega_2(\gamma_0, n, \gamma_k)}{E_{\gamma_k} - E_{n}} , 
\end{eqnarray}
\newred{where the degenerate subspace is given by  $\mathfrak{D} =  \left\{  \ket{+1/2, 1,\up}, \ket{-1/2, 1,\down} \right\}$. Here, we have} defined the product of the matrix elements as 

\begin{eqnarray}
\Omega_1(\gamma_0; n,s; \gamma_k)  = \bra{\gamma_0} \varH_{\mrm{R}}  \ket{n,s} \bra{n,s} \varH_{e-ph} \ket{\gamma_k},   
\end{eqnarray}

\begin{eqnarray}
\Omega_2(\gamma_0; n,s; \gamma_k)  = \bra{\gamma_0} \varH_{\mrm{e-ph}} \ket{n,s} \bra{n,s} \varH_{\mrm{R}}  \ket{\gamma_k} . 
\end{eqnarray}
The matrix elements of the Rashba SO coupling give the selection rule $\left| j - j^{\prime}   \right| = 1$ \cite{refstruck}. 
These transitions are compatible with the selection rules of the electron-phonon interaction mechanisms depending on the order of the dipole expansion considered in the term $e^{\pm i \mathbf{q} \cdot \mathbf{r}}$ \cite{Struck10}.
In this instance, the selection rules match $\left| j - j^{\prime}   \right| = 1$ for the first order and zero order of the dipole expansion of the deformation potential (LA) and bond-length change (LA, TA), respectively.  
For the intrinsic SO, the matrix element of the spin-phonon Hamiltonian is given by $\bra{n_0, \down} \varH_{\mrm{s-ph}}^{i} \ket{n_0, \up} = \bra{n_0, \down} \left[ \mathcal{S}^\mrm{i},  \varH_{\mrm{e-ph}}  \right] \ket{n_0, \up}$, with the ground state set of angular and radial quantum numbers $n_0=(1/2,1)$, since $\varH_{\mrm{i}}$ does not connect the quantum states related with the crossed energy levels. 
Explicitly, we have

\begin{equation}
\bra{n_0, \down} \varH_{\mrm{s-ph}}^{i} \ket{n_0, \up} \propto  \sum_{n^{\prime} \neq n_0}  \delta_{j,j^{\prime}}  \left(   N^{AA}_{n_0 n^{\prime}}  -  N^{BB}_{n_0 n^{\prime}}   \right),
\end{equation}
where $N^{AA}_{nn^{\prime}}= \int \mathrm{dr} \; r \chi^{n}_A(r) \chi^{n^{\prime}}_A(r)$ and $N^{BB}_{nn^{\prime}}= \int \mathrm{dr} \; r \chi^{n}_B(r) \chi^{n^{\prime}}_B(r)$. The selection rule of the intrinsic SO is $\left| j - j^{\prime}   \right| = 0$ which is compatible with the the zero order and first order of the dipole expansion of the deformation potential (LA) and bond-length change (LA, TA), respectively.
The functions $\chi^{n}_A(r)$ and $\chi^{n}_B(r)$ are respectively, purely real and purely imaginary. Thus $\varH_{\mrm{s-ph}}^{i}$ can be rewritten as proportional to $\braket{j,\upsilon}{j,\upsilon^{\prime}}$ with $\upsilon \neq \upsilon^{\prime}$ which is identically zero.
Consequently, the admixture mechanism due to the intrinsic SO does not contribute to the spin relaxation and dephasing process within our model. 


In addition to the admixture mechanism, the spin relaxation can also take place due to the direct coupling of spin and local out-of-plane deformations of the graphene sheet (deflection coupling mechanism)\cite{Rudner10, Struck10}. 
Assuming small amplitudes for the displacement compared to the phonon wavelength, the normal vector to the graphene sheet is $\hat{n}(z) \approx 	\hat{z} + \nabla u(x,y)$. The displacement operator is given by $u_z = \sqrt{1/ A \rho \omega_q}  (e^{i {\bf q} {\bf r}} b^{\dagger} -  e^{-i {\bf q} {\bf r}} b   )$, where we consider linear and quadratic behaviors to the dispersion relation $\hbar \omega_q = \hbar s q + \hbar \mu q^2$, where $\mu = \sqrt{\kappa/\rho}$, with the \new{bending} rigidity $\kappa= 1.1$ eV. 
The matrix element of the effective Hamiltonian containing only the terms connecting the Zeeman levels of the ground state reads

\begin{eqnarray}
\bra{n_0, \down} \varH_{\mrm{s-ph}}^{ZA}  \ket{n_0, \up} &=& \frac{i \lambda_{\mrm{i}} }{\sqrt{A \rho \omega_q}}  \left(  q_x  + i q_y    \right) \\ \nonumber
  &&   \times \left(   N^{AA}_{n_0 n_0}  +  N^{BB}_{n_0 n_0}   \right) , 
\label{eq:hsphflex}
\end{eqnarray} 
where $s_{ZA}=0.25 \times 10^3 \mrm{m/s}$ is the sound velocity.
Here, only the lowest order of the dipole approximation gives a nonzero contribution. 
The spin-phonon terms presented here will be used to calculate the spin relaxation and dephasing rates in the following sections. 

\section{\label{sec:spinrelaxationrates} Spin relaxation rates}

In this section, we calculate the spin relaxation time using the effective spin-phonon Hamiltonian derived in the previous section. 
First, we introduce the Bloch-Redfield theory\cite{blumsbook, chirollisreview}, which allows us to derive the general expression for the spin relaxation and decoherence times.
Consider a general Hamiltonian given by $\varH = \varH_{S} + \varH_{B} + \varH_{SB}$, where $\varH_{S}$ describes the system, 
$\varH_{B}$ a reservoir in thermal equilibrium (bath) 
and $\varH_{SB}$ describes the interaction between them.
This general Hamiltonian $\varH$ is analogous to the one derived in Sec.~\ref{sec:effsph} for all electron-phonon mechanisms and SO interactions via the mapping, 
$\varH_{S} \rightarrow \varH_{\mrm{d}} + \varH_{\mrm{Z}}$, $\varH_{B} \rightarrow \varH_{\mrm{ph}}$ and $\varH_{SB} \rightarrow \varH_{\mrm{s-ph}}$.
The system and the bath are uncorrelated initially, i.e., their spin matrices $\rho$ can be separated as $\rho(0) = \rho_{S}(0) \rho_{B}(0)$. Nevertheless, as time goes by, the system and the bath become correlated via the interaction term $\varH_{\mrm{s-ph}}$. 
This system dynamics is described by an equation of motion for the density matrix in the interaction picture ($\hat{\rho} = e^{i(\varH_{\mrm{d}} + \varH_{\mrm{Z}} + \varH_{\mrm{ph}} )t/\hbar} \rho e^{-i(\varH_{\mrm{d}} + \varH_{\mrm{Z}} + \varH_{\mrm{ph}} )t/\hbar}$) with the bath variables traced out $\hat{\rho}_{S} = \mrm{Tr}_{B} \left[ \hat{\rho} \right]$ as

\begin{equation}
\frac{d}{dt}\hat{\rho}_{S}(t)=  - \frac{i}{\hbar} \int_{0}^{t} d t^{\prime} \mrm{Tr}_{B} \left[ \hat{\varH}_{\mrm{s-ph}}(t), \left[ \hat{\varH}_{\mrm{s-ph}}(t^{\prime}), \hat{\rho}_{S}(t^{\prime}) \hat{\rho}_{B}(0)  \right]\right] 
\label{eq:rho}
\end{equation}
This equation of motion for the reduced density matrix is called the Nakajima-Zwanzig equation\cite{chirollisreview}.
If we assume that the coupling system-bath is weak, this equation can be further simplified by neglecting terms up to $\mathcal{O} (\varH_{\mrm{\mrm{s-ph}}}^2)$ in Eq.~(\ref{eq:rho}), which is equivalent to approximating the density matrix in the integral as $\rho(t) = \rho_S(t) \rho_B (0) + \mathcal{O} (\varH_{\mrm{\mrm{s-ph}}})$ (Born approximation).
Considering a phonon bath, we assume that the time evolution of the $\rho_{S}(t)$ depends only on its present value and not on its past state (Markov approximation), i.e.,  $\hat{\rho}(t^{\prime}) \rightarrow \hat{\rho}(t)$ in the integral of Eq.~(\ref{eq:rho}).
Taking the matrix elements of Eq.~(\ref{eq:rho}) between the eigenstates of $\varH_{S}$, we have that
\begin{eqnarray}
\frac{d}{dt}\hat{\rho}_{S m n}(t)&=& -\frac{i}{\hbar} \omega_{m n}\rho_{m n} - \sum_{k,l} R_{nmkl} \rho_{kl}(t) 
\label{eq:rho2}
\end{eqnarray}
where $\rho_{m n}=\bra{m} \rho \ket{n}$ and $\omega_{nm} = \omega_n - \omega_m$. The term $R_{nmkl}$ is the Redfield tensor 

\begin{eqnarray}
R_{nmkl} = \delta_{nm} \sum_r \Gamma_{nrrk}^{+} + \delta_{nk} \sum_r \Gamma_{lrrm}^{-} - \Gamma_{lmnk}^{+} - \Gamma_{lmnk}^{-}, 
\end{eqnarray}
where $\Gamma^{+}_{\mrm{lmnk}}= \int_{0}^{\infty} \mrm{dt} e^{- i \omega_{nk} t} \overline{ \bra{l} \varH_{\mrm{s-ph}} \ket{m}  \bra{n} \varH_{\mrm{s-ph}}(t) \ket{k}  }$, with $ \Gamma^{+}_{lmnk} = \left(  \Gamma^{-}_{knml} \right)^{*}  $. Here, the overbar denotes the average over a phonon bath in thermal equilibrium at temperature $T$. 
Using Eq.~(\ref{eq:rho2}) in the secular approximation where $R_{nmkl}$ is approximatedly given by a diagonal tensor and $\left\langle  \mrm{d S_z/dt} \right\rangle = \mrm{Tr}[(\mrm{d\rho/dt)} S]$, we can derive the differential equation describing time evolution of the average values of the spin components, also known as Bloch equations. 
The solution for the $\left\langle  S_z \right\rangle$ component with a magnetic field applied along the same direction is 
$\left\langle  S_z \right\rangle (t) = S_z^0 - (S_z^0 - S_z(0))e^{-t/T_1}$, where $S_z^0$ is the equilibrium spin polarization (ensemble of spin-down electrons)	and $S_z(0)$ is the initial non-equilibrium spin alignment considered in the problem (ensemble of spin-up electrons).  

Explicitly, the spin relaxation rate is given by\cite{chirollisreview}

\begin{equation}
\Gamma_{\down \up}=\frac{1}{T_1} = 2 \mathfrak{R} \left(  \Gamma^{+}_{\gamma_0 \gamma_k \gamma_k \gamma_0} +  \Gamma^{+}_{\gamma_k \gamma_0 \gamma_0 \gamma_k}   \right)  ,
\label{eq:t1}
\end{equation}
Equation ~(\ref{eq:t1}) can be simplified to 

\begin{equation}
\frac{1}{T_1} = \frac{2 \pi}{\hbar} \sum_q \left|  \bra{\gamma_0} \varH_{\mrm{s-ph}} \ket{\gamma_{k}}  \right|^2 \delta(\hbar\omega_{\gamma_0 \gamma_k} - \hbar\omega_q) \mrm{coth} \left(  \frac{\hbar \omega_{\gamma_0 \gamma_k}}{2 k_b T} \right) .
\label{eq:t1goldenrule}
\end{equation}

The spin relaxation rate is then calculated combining Eqs.~(\ref{eq:t1goldenrule}) and ~(\ref{eq:hsphrashba}).
The contribution due to the deformation potential (LA) combined with the Rashba SO coupling is given by

\begin{eqnarray}
\Gamma_{\gamma_0 \leftarrow \gamma_k}^{g1:LA} = \frac{\pi}{2} \frac{g_1^2}{\hbar \rho s_{LA}^2} \left( \frac{ E_{\gamma_k} - E_{\gamma_0}}{ \hbar  s_{LA}}    \right)^4 
\int_0^{2 \pi} \mrm{d}\phi_q \left[  \Lambda_i^k (A_{g_1})   \right]^2 .
\end{eqnarray}
And those due to the bond-length change mechanism for $\mu=\mrm{LA}, \mrm{TA}$, 

\begin{eqnarray}
\Gamma_{\gamma_0 \leftarrow \gamma_k}^{g2:LA,TA} = 2 \pi \frac{g_2^2}{\hbar \rho s_{\mu}^2} \left( \frac{ E_{\gamma_k} - E_{\gamma_0}}{ \hbar  s_{\mu}}    \right)^2 
\int_0^{2 \pi} \mrm{d}\phi_q \left[  \Lambda_i^k (A_{g_2})   \right]^2 ,
\end{eqnarray}
where we imply summation over the repeated index $i=1,2,3$. 
In the above we have define  

\begin{eqnarray}
\Lambda_1^{k}(A_{g_1, g_2}) = \lambda_1^{n} \bra{1/2,1, \down}  A_{g_1, g_2} \ket{-1/2,1, \up} \rho_k, 
\end{eqnarray}

\begin{eqnarray}
\Lambda_2^{k}(A_{g_1, g_2}) =  \sum_{n \neq (1/2,1)}  \lambda_2^{n} \bra{1/2,1, \down}  A_{g_1, g_2} \ket{n, \down} \\ \nonumber
\times \bra{n, \down}  H_R \ket{1/2,1, \up} \sigma_k, 
\end{eqnarray}

\begin{eqnarray}
\Lambda_3^{k}(A_{g_1, g_2}) = \sum_{n \neq (1/2,1)} \lambda_3^{n} \bra{1/2,1, \down} H_R \ket{n, \up} \\ \nonumber
\times \bra{n, \up}  A_{g_1, g_2} \ket{1/2,1, \up} \sigma_k, 
\end{eqnarray}
where $A_{g_1} = a_1 \openone_{2x2}$, $A_{g_2} = g_2 \left(  \sigma_+ a_2^{*} + \sigma_- a_2   \right)$, with $\sigma_{\pm} = (\sigma_x \pm i \sigma_y)/2$. 
Their respective matrix elements are given by

\begin{eqnarray}
\bra{n} A_{g1} \ket{n^{\prime}}&=&  M_{nn^{\prime}}
 \left(  \delta_{j,j^{\prime}+1}e^{-i \phi_q}  + \delta_{j,j^{\prime}-1}e^{+i \phi_q}    \right)  ,
\end{eqnarray}
with $M_{n n'}=\int \mrm{dr}\; r^2 \left( {\chi_{A}^{n}}^* \chi_{A}^{n '}
+{\chi_{B}^{n}}^*\chi_{B}^{n'}\right)$, and 

\begin{eqnarray}
\bra{n} A_{g2} \ket{n^{\prime}}&=&     \left(   g_2 a_2^{*} \delta_{j,j^{\prime}+1}N^{AB}_{nn^{\prime}} +   g_2 a_2 \delta_{j,j^{\prime}-1}N^{AB}_{n^{\prime}n} \right),  
\end{eqnarray}
where $N^{AB}_{nn^{\prime}}= \int \mathrm{dr} \; r \chi^{n}_A(r) \chi^{n^{\prime}}_B(r)$. 
Here, $\rho_{\gamma_1} = -\sin(\vartheta/2)$, $\sigma_{\gamma_1} = \cos(\vartheta/2)$ and $\rho_{\gamma_2} = \cos(\vartheta/2)$, $\sigma_{\gamma_2} = \sin(\vartheta/2)$.  
The energy-dependent denominators are given by $\lambda_1^{n} = 1$, $\lambda_2^{n} = 1 / E_k - E_{n} + g\mu_B B/2$, $\lambda_3^{n} = 1/ E_{1/2,1} - E_{n} - g\mu_B B/2$. 

\begin{figure}[htb]
\includegraphics[width=.48\textwidth]{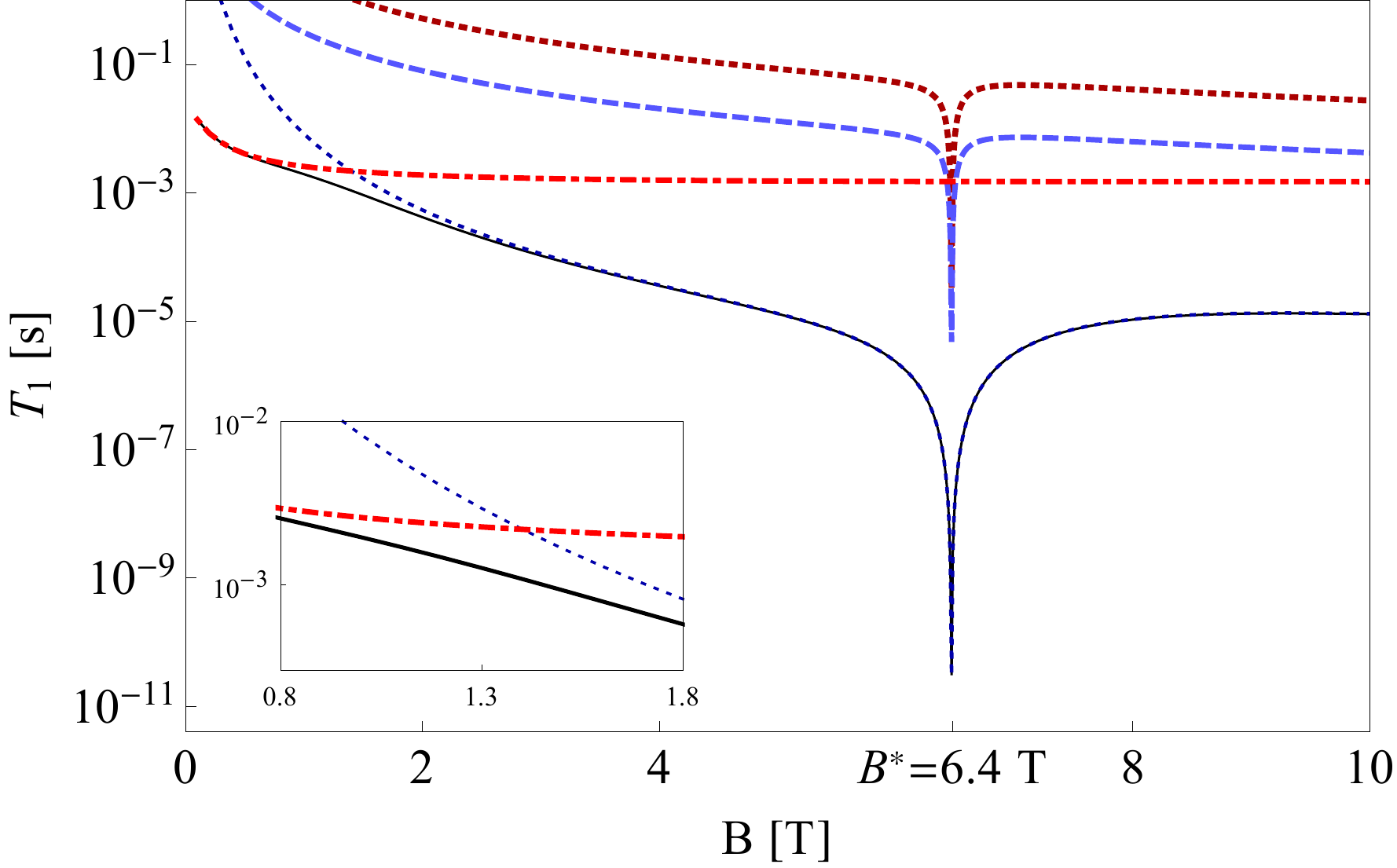}
\caption{\label{fig:plot2graphene} Magnetic field dependence of the spin relaxation time. 
Parameters used in the numerical evaluation are given in Table~\ref{tab:table2}.
Contributions from the deformation potential $\mrm{g_1:LA}$ (dark, dotted), bond-length change mechanism $\mrm{g_2:LA}$ (dark, dotted) and $\mrm{g_2:TA}$ (light, dashed) and the out-of-plane phonons $\mrm{ZA}$ (light, dot-dashed). 
Dark solid: the sum of all processes.
The minimum in $T_1$ occurs at the energy-level anticrossing at $B^{*}$.
Inset: Blowup of the low magnetic field regime. Competition between the two electron-phonon dominant mechanisms: deformation potential and flexural phonons. The absence of Van Vleck cancellation\cite{VanVleck40, Struck10} leads to a finite value for $T_1$ at $B=0$.}
\end{figure}

As stated in Sec.~\ref{sec:deg}, the energy relaxation accompanied by a spin-flip transition occurs between the states $\ket{\gamma_0}$ and $\ket{\gamma_1}$ before the energy anticrossing $\Gamma_{\down \up} = \Gamma_{\gamma_0 \leftarrow \gamma_1}$, and between the states $\ket{\gamma_0}$ and $\ket{\gamma_2}$ after the energy anticrossing $\Gamma_{\down \up} = \Gamma_{\gamma_0 \leftarrow \gamma_2}$, for all electron-phonon mechanisms $\Gamma_{\down \up}^{R} = \Gamma_{\gamma_0 \leftarrow \gamma_k}^{g1:LA} + \Gamma_{\gamma_0 \leftarrow \gamma_k}^{g2:LA} + \Gamma_{\gamma_0 \leftarrow \gamma_k}^{g2:TA}$. 
The contribution from the out-of-plane flexural phonons via the deflection coupling mechanism, calculated using Eq.~(\ref{eq:t1goldenrule}) combined with Eq.~(\ref{eq:hsphflex}), is
\begin{eqnarray}
\Gamma_{\down  \up}^{ZA} = \frac{4 \pi^2}{\rho} \frac{\lambda_i^2}{g \mu_B B} \frac{1}{Q(B)} \left(  \frac{-s_{ZA} + Q(B)}{2 \mu}   \right)^3  \\ \nonumber 
													\times 	\left|	\int\mrm{d}r\; r \left(	\left| \chi^{n}_{A}\right|^2 - \left| \chi^{n}_{B}\right|^2	\right) \right|^2   ,
\end{eqnarray}
where we define $Q(B) = \sqrt{s_{ZA}^2 + 4 \mu (g \mu_B B/\hbar)}$, with $s_{ZA} = 0.25 \times 10^3 \mrm{m/s}$. 
\newred{In the low magnetic field limit, the term $\Gamma_{\down  \up}^{ZA}$ simplifies to}
\begin{equation}
\Gamma_{\down  \up}^{ZA} = \frac{4 \pi^2\lambda_i^2}{\rho}\frac{1}{s_{ZA}^5} \left( g \mu_B B \right)^2
													\times 	\left|	\int\mrm{d}r\; r \left(	\left| \chi^{n}_{A}\right|^2 - \left| \chi^{n}_{B}\right|^2	\right) \right|^2 .
\end{equation}
The magnetic field dependence of $T_1 = (\Gamma_{\down \up}^{R} + \Gamma_{\down \up}^{ZA})^{-1}$ with all the mechanisms considered in this work is evaluated numerically and is presented in Fig.~\ref{fig:plot2graphene}. 
It can be observed that at the energy anticrossing region, the spin relaxation time rapidly decreases, characterizing its non-monotonic behavior induced by an external electric field via the Rashba SO interaction. Notice that if no external electric field is applied, the spin relaxation time is monotonic with contributions from only the intrinsic SO interaction	via deflection coupling mechanism. 
\begin{table}[b]
\caption{\label{tab:table2}
Parameters for the numerical evaluation of the spin relaxation rates. The electron-phonon coupling constants for the deformation potential $g_1$ and for the bond-length change mechanism $g_2$ and the coupling strengths for Rashba $\lambda_R$ for an external electric field $E$ and the intrinsic $\lambda_i$ SO couplings. 
The graphene layer is characterized by its mass area density $\rho$. 
The quantum dot parameters are its radius $R$, potential height $U_0$ and the substrate-induced energy gap $\Delta$. The system is assumed to be in thermal equilibrium with the bath at temperature $T$.
}
\begin{ruledtabular}
\begin{tabular}{cccc}
$g_1$ & $30$ eV\footnotemark[1]  \\
$g_2$ & $1.5$ eV\footnotemark[1] \\
$\lambda_R$ & $11$ $\mu$eV\footnotemark[2] \\
$E$ & $50$ V/$300$ nm\footnotemark[3] \\
$\lambda_i$ & $12$ $\mu$eV\footnotemark[4] \\
$\rho$ & $7.5 \times 10^{-7}$ $\mrm{kg/m^2}$\footnotemark[5] \\
$R$ & $35$ $nm$ \\
$U_{0} = \Delta$ & $260$ meV \\
$T$  & $100$ mK \\
\end{tabular}
\end{ruledtabular}
\footnotetext[1]{From Ref.~\onlinecite{Ando05}.}
\footnotetext[3]{From Ref.~\onlinecite{KaneMele05}.}
\footnotetext[3]{From Ref.~\onlinecite{Min06}.}
\footnotetext[4]{From Ref.~\onlinecite{Gmitra09}.}
\footnotetext[5]{From Ref.~\onlinecite{Falko08}.}
\end{table}

%

%
The magnetic field dependence of the spin relaxation rate for each electron-phonon coupling mechanism can be understood using the spectral density of the system-bath interaction

\begin{eqnarray}
J_{\gamma_0 \gamma_k}(\omega)= \int_{-\infty}^{\infty} \mrm{dt} e^{- i \omega t} \overline{ \bra{\gamma_0} \varH_{\mrm{s-ph}}(0) \ket{\gamma_k}  \bra{\gamma_k} \varH_{\mrm{s-ph}}(t) \ket{\gamma_0}  }. 
\end{eqnarray}
Further simplifications in Eq.~\ref{eq:t1} allow us to find the following relation
\new{$ 1/T_1  \propto J_{\gamma_0 \gamma_k}(\omega_{\gamma_0 \gamma_k})$}, 
where $\omega_{\gamma_0 \gamma_k}
\propto  \omega_Z \propto g \mu_B B$. In a general form, we have that 
$1/T_1  \propto  \sum_q  K_q \bra{\gamma_0} e^{i \mathbf{q} \cdot \mathbf{r}} \ket{\gamma_k}  \bra{\gamma_k} \varH_{\mrm{SO}} \ket{\gamma_0} \delta(\omega_q - \omega_{\gamma_0 \gamma_k})$, where $K_q = q / \sqrt{\omega_q}$ since $\varH_{e-ph} \propto K_q e^{\pm i \mathbf{q} \cdot \mathbf{r}}$. Also, $\sum_q \propto \int \mrm{dq} \; q^{d-1}$, where $d=2$ is the dimensionality of graphene.
Each SO coupling defines the selection rule for the quantum number $j$ and consequently, the order of the dipole expansion as explained in Sec.~\ref{sec:effsph}.
We find that for the Rashba SO coupling, \new{$J_{\gamma_0    \gamma_k}(\omega_{\gamma_0 \gamma_k}) \propto \omega_Z^s$}
with $s=4$ for the deformation potential (LA) and $s=2$ for the bond-length change mechanism (LA, TA).  
Also, for the intrinsic SO, $s \geq 2$ for the direct spin-phonon coupling (ZA). Therefore the spectral density of the system-bath interaction is super-Ohmic ($s>1$) with a strong dependence with the bath frequency for all phonons considered in graphene.

\section{\label{sec:spindephasingrates} Spin dephasing rates}

Next we evaluate the spin dephasing rates for all the electron-phonon mechanisms introduced in Sec.~\ref{sec:effsph}.  
Within the Bloch-Redfield theory, we can also solve the Bloch equations for the spin components perpendicular to the magnetic field, which are given by $\left\langle  S_x \right\rangle (t) = S_{x}^0 \cos(\omega_Z t) e^{-t/T_2}$ and $\left\langle  S_y \right\rangle (t) = S_{y}^0 \sin(\omega_Z t) e^{-t/T_2}$, where $S_{x,y}^0$ are the initial spin polarizations along the $x,y$ directions.
The decoherence time can be separated into two contributions: the spin relaxation and the pure spin dephasing $1/T_2 = 1/2 T_1 + 1/T_{\phi} $, where the pure spin dephasing rate is\cite{chirollisreview} 

\begin{equation}
\Gamma_{\phi}=\frac{1}{T_{\phi}} = \mathfrak{R} \left( \Gamma_{\gamma_0 \gamma_0 \gamma_0 \gamma_0}^{+} + \Gamma_{\gamma_k \gamma_k \gamma_k \gamma_k}^{+} - 2 \Gamma_{\gamma_0 \gamma_0 \gamma_k \gamma_k}^{+}     \right).
\label{eq:tphi}
\end{equation}
In the low-temperature limit, we find that 
\begin{equation}
\frac{1}{T_{\phi}} = 	\lim_{\omega \to  0} \left| \bra{\gamma_0} \varH_{\mrm{s-ph}} \ket{\gamma_0} -  \bra{\gamma_k} \varH_{\mrm{s-ph}} \ket{\gamma_k} \right|^2 \delta(\hbar \omega - \hbar \omega_q) \frac{2\pi k_b T}{\hbar \omega}. 
\label{eq:tphi2}
\end{equation}
The dephasing time can also be rewritten in terms of the spectral density of the system-bath interaction as $ 1/T_{\phi}  \propto \lim_{\omega \rightarrow 0} J(\omega) \mrm{coth} \left(  \hbar \omega/2 k_b T \right)  \propto  \lim_{\omega \rightarrow 0}  J(\omega) / \omega $.  
As we have \new{shown} in Sec.~\ref{sec:effsph}, the spectral function for all electron-phonon coupling mechanisms considered in this work are super-Ohmic. Thus the spin dephasing vanishes in all cases, since  $ 1/T_{\phi}  \propto \lim_{\omega \rightarrow 0}  \omega^s / \omega \rightarrow 0$, with $s >1$. 
In \new{other} words, there are no \new{phonons} available in leading order to cause dephasing in graphene quantum dots. 
The decoherence time $T_2$ is determined only by the
relaxation contribution, i.e., $T_2 = 2 T_1$. Notice that this
relation is no longer necessarily true considering two-phonon
processes since the combination of emission and absorption energies
can fulfill the energy conservation requirement.\new{\cite{Golovach04}}

Aditionally, the spin dephasing rate \new{could also} vanish at the energy anticrossing for a super-Ohmic bath. 
Within the subspace spanned by the states $\left\{  \ket{+1/2, 1,\up}, \ket{-1/2, 1,\down} \right\}$, the Hamiltonian $\varH$ can be rewritten as $\varH_{\phi} = \Delta_+(B) 1 + \Delta_-(B) \tau_z$, where $\tau_z$ denote a Pauli matrix  and $\Delta_{\pm} = (E_{\gamma_3} \pm E_{\gamma_2})/2$. This magnetic field can be divided into two contributions $B = B_{0} + \delta B(t)$: an external source $B_{0}$ and an internal contribution $\delta B(t)$ due to the bath. For small fluctuation of $\delta B(t)$, the $\varH_{\phi}$ is approximatedly given by

\begin{equation}
\varH_{\phi}=  (\Delta_-(B_0) +  \partial_{B} \Delta_-(B_0) \delta B(t) ) \tau_z    ,
\label{eq:Hphi} 
\end{equation}
where we have not included the term proportional to $\Delta_+ \openone_{2x2}$ since it does not cause spin dephasing. 
Calculating the spin dephasing rate within the Bloch-Redfield theory using Eq.~\ref{eq:tphi}, we find that 

\begin{equation}
\frac{1}{T_{\phi}}=  \left( \frac{2}{\hbar}  \partial_{B} \Delta_-(B_0) \right)^2 \lim_{\omega\to  0} \mathfrak{R} \int_{0}^{\infty} \mrm{dt}^{\prime} e^{-i \omega t} \overline{\left\langle \delta B(0) \delta B(t^{\prime})    \right\rangle}   ,
\label{eq:tphi3}
\end{equation}
where $\left\langle A(t) \right\rangle$ is the thermal equilibrium expectation value of the operator $A(t)$ on the bath. Therefore the spin dephasing rate goes to zero at the energy anticrossing, since $ \partial_{B} \Delta_-(B_0) \rightarrow 0$. 
This condition is valid under the assumption that the thermal average of the fluctuating magnetic field does not diverge.   
Following the result given by Eq.~(\ref{eq:tphi2}), the spin dephasing rate still vanishes as long as the spectral density of the system-bath interaction is super-Ohmic, i.e., $J(\omega) \propto \omega^s$, with $s>1$.

\section{\label{sec:conclusion} Conclusion}

In summary, we find a minimum in the spin relaxation time as a function of the magnetic field  that is induced by the Rashba SO coupling and is controllable by an external electric field. 
In larger quantum dots, the intrinsic SO dominates the spin relaxation over the Rashba SO contribution at low magnetic fields. As the magnetic field increases, the extrinsic contribution takes over, generating a non-monotonic behaviour of $T_1$ due to the Rashba SO interaction-induced level anticrossing.
We have also analyzed the spectral density of the system-bath interaction for the first-order electron-phonon interaction and we have identified a vanishing contribution to the energy-conserving dephasing process. 
Therefore the phonon-induced pure spin dephasing rate is of the same order of magnitude as the spin relaxation rate, i.e., $T_2 = 2 T_1$, in the leading order of the electron-phonon interaction.
Other mechanisms such as nuclear spins from the $^{13}C$ atoms and charge noise combined with SO interaction could lead to a non-vanishing spin dephasing rate. Nevertheless, these mechanism are expected to be weak in graphene\cite{Guido07, Fischer09}. 
Moreover, we have shown that any super-Ohmic bath has a vanishing spin dephasing rate at the energy anticrossing.

\begin{acknowledgments}
We wish to acknowledge useful discussions with P. R. Struck and Peter Stano. 
\new{Funding for this work was provided by CNPq, FAPESP and PRP/USP within the Research Support Center initiative (NAP Q-NANO) (MOH and JCE) and DFG and ESF under grants FOR912, SPP1285, and EuroGraphene (CONGRAN) (GB). }
\end{acknowledgments}


\begin{thebibliography}{99}
\bibitem{tensor}
The tensor product notation is implicit, i.e., $\sigma_i s_j \equiv \sigma_i \otimes s_j$, for $i,j=x,y,z$.
\bibitem{sphonon}
We do not carry the term $\left[  \mathcal{S}, \varH_{ph}  \right]$ since it does not contribute to the effective spin-phonon Hamiltonian, i.e., $\bra{\gamma_0} \varH_{ph}+ \left[  \mathcal{S}, \varH_{ph}  \right] \ket{\gamma_k} = 0$.
\bibitem{refstruck}
	The matrix elements of the Rashba SO and all the electron-phonon coupling mechanisms are presented in Ref.~[\onlinecite{Struck10}].
\end{thebibliography}
\end{document}